\documentclass[letterpaper,twocolumn,10pt]{article}
\usepackage{usenix2019,epsfig,endnotes}
\usepackage{array}
\usepackage{algorithm}
\usepackage[noend]{algpseudocode}
\usepackage{algorithmicx}
\usepackage{comment}
\usepackage{adjustbox}
\usepackage{xcolor}
\usepackage{multirow}
\usepackage{float}
\begin{document}

%don't want date printed
\date{}

\title{\Large \bf Pilot evaluation of Collection API with PID Kernel Information}

\author{
{\rm Yu Luo}\\
School of Informatics, Computing, and\\ Engineering\\
Indiana University\\
Email: luoyu@indiana.edu
\and
{\rm Beth Plale}\\
School of Informatics, Computing, and\\ Engineering\\
Indiana University\\
Email: plale@indiana.edu
} % end author

\maketitle

\thispagestyle{empty}

\subsection*{Abstract}
As digital data become increasingly available for research, there is growing awareness of the value of domain agnostic Persistent Identifiers (PIDs) for data.  A PID is a globally unique reference to a digital object, which in our case is data. In an ecosystem of connected digital objects, a PID will reference a digital object, and the digital object will be a simple entity, a collection of homogeneous objects, or a set of heterogeneous objects.

In this paper we study two recent recommendations from the Research Data Alliance (RDA) that both address pieces of an ecosystem of connected digital objects.  The recommendations address Persistent ID records and representations of collections of data.  We evaluate different approaches in where to locate key information about a data collection between these two component solutions.   

\section{Introduction}

The Research Data Alliance (RDA)~\footnote{rd-alliance.org} has fostered the development of component parts of an ecosystem of connected digital objects, wherein all research data take the form of digital objects and all digital objects have a globally unique persistent ID.  This model has its roots in Kahn's Digital Object Architecture~\cite{kahnobjectmodel}.  Several RDA working groups (WG) that contribute to the ecosystem of connected digital objects include the Data Type Registry WG~\cite{dataTypeRegistry}, the Persistent Identifier Kernel Information (PID KI) WG~\cite{PIDKI-zenodo}, and Collection API WG~\cite{collection:api}.  The Data Type Registry is envisioned as a globally accessible registry for type information that enables the interpretation of digital objects (for example, providing information that can answer the question "what tool can I invoke to read this object?")  The PID KI WG proposes a small amount of information embedded in a Persistent Identifier.  The RDA working group for PID KI undertook to define the general principles of PID Kernel Information that can be stored as a set of $\{name, value\}$ pairs. Finally, the Collection API working group defined a minimal interface for accessing and interpreting digital objects that take the form of collections of objects.  

Drawing from our earlier work in data provenance~\cite{suriarachchi2015komadu}, we posit that the provenance of a digital object can aid in its findability and accessibility. This occurs when all digital data are digital objects with Handles - a well known persistent ID.  Is one digital object a revision of another or is it a version~\cite{trustthreads}?  If it is a version, then it is important to ensure that the only person asserting the version relationship is the owner of the original digital object. This constraint may not hold for revisions.  

Drawing on the terminology of W3C PROV \footnote{https://www.w3.org/TR/prov-overview/}, we identify a set of object-to-object provenance relationships that are candidates for inclusion in the PID Kernel Information record. This provenance we refer to as "backbone provenance" as it is central to linking between digital objects.  We describe backbone provenance in more detail in Section~\ref{sec:problemdefinition}.  

There is an issue with representing a collection as PID Kernel Information, however, as a collection can be arbitrarily deep or wide.  As the information included in a PID record needs to be limited in size, a data structure of unknown size is not ideal. Hence we explored the Collection API as a solution for representing digital objects that are related to one another through a membership relationship. 

The Collection API as defined by ~\cite{collection:api}, defines an application programming interface (API). We use a prototype implementation of the API developed by Tufts University.  The Collection API defines a fine-grained collection object as a structure of \textit{Collections} and \textit{Members}. The API supports reconstructing the structure of objects.

This paper compares the tradeoffs in distribution of metadata, particularly data provenance, across the PID KI record and the Collection API.  Under what conditions is it better to consult the Collections API to resolve the linkings between related digital objects?   The practicality of embedding collection information in the PID Kernel Information has strong advantages and disadvantages, suggesting that an alternate approach is preferred in some cases.  

The remainder of this paper is organized as follows. Section~\ref{sec:relatedWork} presents the related work while Section~\ref{sec:problemdefinition} defines the problem space, Handle system, Collection API and PID Kernel Information. Section~\ref{sec:mapping} discusses the two data models:  PID KI and Collections API. Section~\ref{sec:usecase} is use cases. Section~\ref{sec:solution} describes the distribution strategy of maintaining provenance in PID KI and Collection API. The experiment architecture and environment of the experiment are presented in Section~\ref{sec:evaluation}. Section~\ref{sec:result} presents the result of PID KI and Collection API study in four use cases and three provenance distribution strategies. Section~\ref{sec:discussion} presents the understandings and explanations of provenance distribution strategies in different use cases. Section~\ref{sec:conclusion} concludes the the work.

\section{Related Work}\label{sec:relatedWork}

The DOI community has examined kernel information records for PIDs. Paskin \cite{paskin2010digital} defines metadata, the DOI Kernel, and the minimum set of metadata information specified at the time a DOI is created. The PID Kernel Information draws parallels to the DOI Kernel; the latter has been discussed but never adopted into the DOI system. 

The Archival Resource Key (ARK) \cite{kunze2003towards} is a persistent ID. The ARK, unlike the URL, PURL, and URN, provides three things: an object, metadata and a provider's commitment. While adding '$?$' to the end of one ARK link, the server returns the human-readable information, for example, the Electronic Resource Citation using Dublin Core Kernel metadata, including who, when, where and what. This information helps users understand the object. It is possible that the PID Kernel Information approach could be built into the ARK resolution system. 

Research into a network of provenance is carried out in Zhou \cite{Zhou:2010:EQM:1807167.1807234}. Their work, ExSPAN, is a framework that achieves the network provenance at Internet-Scale. The ExSPAN stores provenance in a relational table, and establishes a granular relationship graph of objects for efficient query and maintain. The idea of keeping network provenance is quite similar to the research of this paper, but ExSPAN is working on the local use and analysis. 

Chen \cite{chen2012visualization} provides a approach to visualize the graph of network provenance captured from streaming data. The research is fully maximizing the benefit of provenance, and use visualization techniques to interact with digital object. The idea of representing the network data by using provenance from Chen's research is a good reference to start our research, however, our research wants to make the the Internet Scale of presenting and sharing data, using the minimal metadata information, keeping fine-grain provenance relationship, by using Persistent Identifiers. All these requirements are challenges for current Network Provenance or Provenance Network representation work, and this is the motivation that we do the pilot evaluation on Handle System and Collection API Service, attempting to contribute to the Publishing Ecosystem.

Komadu \cite{suriarachchi2015komadu} is a provenance capture and representation system.  It represents the provenance among different objects and workflows but does so using a centralized approach which limits data sharing on the Internet.

\section{Environment}\label{sec:problemdefinition}

This study examines multiple options to distribute information across PID Kernel Information and the Collection API to maintain the lineage of digital objects.  There are several key concepts and services that we describe here that are the objects of our study.  

Each digital object is identified by a PID. It can be complex on a number of dimensions\cite{DBLP:journals/corr/abs-0707-1534}: \textit{multi-format, multi-structure, multi-source, multi-model} and \textit{multi-version}.  A digital object in the Digital Object Architecture~\cite{kahnobjectmodel} is discrete - that is, it cannot be decomposed, but is related to other digital digital objects.  

\paragraph{Handle System.}
The Handle System~\cite{Sun:2003:HSO:RFC3650} is a global resolution system for persistent IDs.  Under the governance of the DONA Foundation~\cite{DONAfoundation}, the Handle system runs hundreds of Global Handle Registries and Local Handle Servers to resolve a PID to the object to which it refers. A Global Handle Registries reside at well known locations, they strip off the prefix of the PID and hand off remainder of the Handle to the  Local Handle Server that maintains the PID record.  

That is, the Global Handle Registry (GHR) manages the authorized Handles and provides the service information of any Local Handle Service (LHS). The naming authority requests are sent directly to the GHR, and GHR sends back the service information. Then client then follows the service information sending the request to responsible LHS.

\paragraph{PID Kernel Information.}
The relationship between digital digital objects is data provenance.  One digital object can be related to another through a number of provenance relationships.  

The Provenance Working Group of World Wide Web Consortium (W3C) has defined and published a number of standards that define provenance information, for instance PROV DM \cite{w3cmodel}. The core structure of PROV is the entity, the activity, and the agent.  Their relationships are one of action: an "activity", invoked by an "agent", used by an "entity".  

From this action vocabulary, we selected a subset that defines relationships between entities (the digital objects).  We call this \textit{backbone provenance}. The fields are:

\begin{itemize}
    \item 
wasDerviedFrom
\item 
wasRevisionOf 
\item 
wasQuotedFrom 
\item 
hasPrimarySource 
\item 
alternateOf
\item 
hasMember
\item 
specializationOf
\end{itemize}

Backbone provenance is a limited set of W3C PROV definitions, but it is sufficient to represent the relationships between entities, especially the objects in collection model.

The PID Kernel Information WG of the Research Data Alliance undertook to define a minimal amount of information that could be stored at a Local Handle Server to facilitate faster discovery.  As Handles contain zero semantic information, the PID Kernel Information record could support  semantic information that can help a client to make high level decisions about PID relevance.  Interpretation of the PID Kernel Information requires a type definition that is stored in a Data Type Registry~\cite{dataTypeRegistry}. 

PID Kernel Information is not intended to fully describe a digital object, the purpose of PID Kernel Information is making highly efficient high level programmatic decisions though inspection of PID records alone.
 
The Local Handle Service controls a restricted namespace that manages Handles under given naming authority. Accepting the requests from client, LHS sends the Handle results (PID Kernel Information) back to client.  

As stated in the introduction, the Collection API is a recommendation of the Research Data Alliance.  A prototype Collection API Service  has been developed as a prototype of the API. We use the Collection API service in this study. We hereafter refer to the prototype as Collection API to distinguish it from the RDA recommendation.

\paragraph{Collection API}. The Collection API service defines a fine-grained collection object structure as \textit{Collections} and \textit{Members}. Collection capabilities, collection properties and collection membership compose to the first level of data hierarchy, describing the full actions and essential metadata of the collection. Membership of a collection builds the graph with the relative Member objects, linking to the relevant member entities. The second but last layer of hierarchy is the Member entity, which is divided into general properties of location, description, data type and ontology. Note that a collection could be the member of another collection, inheriting the logic of tree structure.

\textit{Collection-Member} architecture could be extended to a hierarchy tree structure, which has infinite possibility on the structure of Digital digital object. The flexibility of the Colllection API service is trying to overcome the challenge of inconsistent structure of digital object before and after processing, which assist Handles (PID KI) in defining the complex structure of digital object.

\section{Collection API and Backbone Provenance }\label{sec:mapping}

A Collection is quite different from a Directory or a Folder. In a file system, the folder or directory bounds a collection of files or other folders or directories. The collection, however, is an object that references other objects (members/collections) through a provenance relationship. 

In a digital object ecosystem, objects reference other objects through the backbone provenance which includes the  'HadMember' relation.   An object that is a member of a collection can also be in a different provenance relationship with other objects in the object ecosystem.

\begin{figure}[h]
\includegraphics[width=0.5\textwidth]{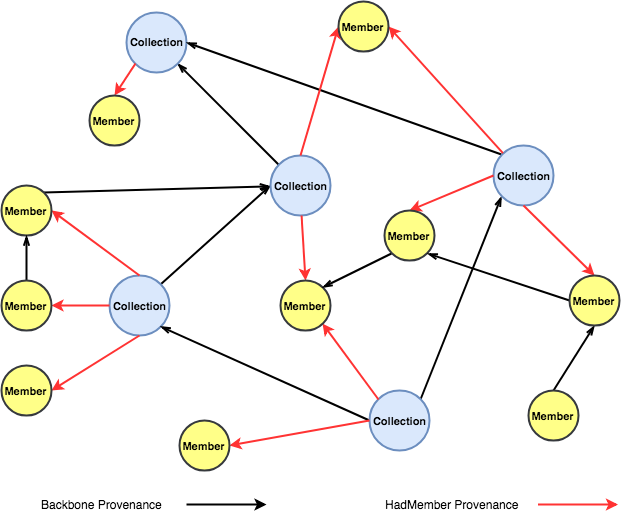}
\caption{Collections exhibiting both membership and non-membership provenance}
\centering
\label{provenancerelationship}
\end{figure}

However, in this research we restrict graphs to be acyclic graphs made up of collections that relate to their member objects through the 'HadMember' relationship, and collections and members that relate to one another through non-hadMember provenance. This more restricted graph can be seen in Fig. \ref{provenancerelationship}

Provenance relations capture activity through time, that is, they exhibit timeliness.  Object $a'$ is a new version of object $a$.  In the provenance relationship, we first publish one object as root, and then published other objects later. The publishing date is a hidden information showing that we cannot find a unpublished data before a published data in Provenance relationship. Directionality indicates the relationship among objects based on timeliness. For example, we could generate a next version object based on the current object, but inverse doesn't work. Directioniality is the visualization of timeliness with actions of the object.

\section{Graph Cases}\label{sec:usecase}

This section defines four synthetic graphs of digital object relationships. A graph is directed and acyclic, of vertices (node) and edges (relationship), $N = \{V, E\}$ where a vertex (node), $V$ corresponds to digital object, edge (relationship), $E$ is a directed acyclic provenance relationship between vertices. 

The four graphs are here and each described in more detail below:
\begin{itemize}
\item
G1. Directed linear backbone relationship among collection objects
\item
G2. Directed acyclic backbone relationship among collection objects
\item
G3. Directed acyclic backbone relationship among collection or member objects
\item
G4. Directed acyclic backbone relationship among all objects
\end{itemize}

 %"HadMember" link is supported by Collection API service, helping to achieve the integration of the complex digital object in data model of PID KI and Collection API object. In these three solutions, Collection API would basically handle "HadMember" provenance relationship for all complex digital objects. Then we gradually transfer the Backbone Provenance information into Collection API object in different structures, testing the collaboration between PID KI and Collection API service on managing and resolving Provenance relationship.

\begin{figure}[h]
\includegraphics[width=0.5\textwidth]{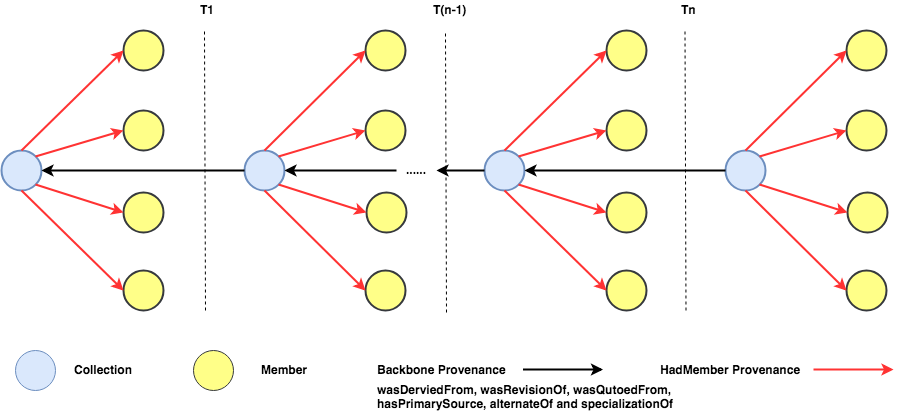}
\caption{Directed linear backbone relationship among collection objects}
\centering
\label{case1}
\end{figure}

\textbf{\textit{G1. Directed linear backbone relationship among collection objects}}.  Linear provenance relationship graph contains collection digital objects linked with Backbone provenance. Each collection object points to 4 member objects with 'HadMember' relationship. The data structure is shown in Fig. \ref{case1} with \textit{\textbf{T(x)}} indicates timeliness and the arrow presents provenance relationship.

\begin{figure}[h]
\includegraphics[width=0.5\textwidth]{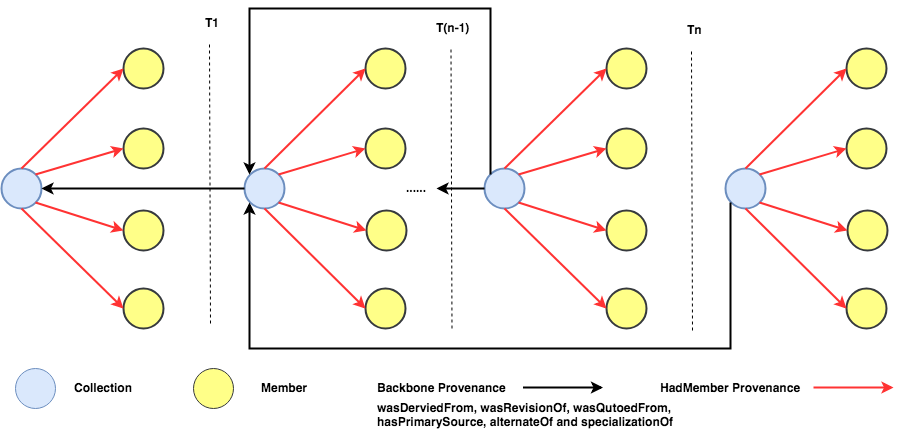}
\caption{Directed Acyclic backbone relationship among collection objects}
\centering
\label{case2}
\end{figure}

\textbf{\textit{G2. Directed Acyclic backbone relationship among collection objects}}.  Directed Acyclic Provenance relationship graph contains collection digital objects linked with Backbone provenance. Each collection object connects 4 member objects by 'HadMember' relationship. In this case, we put more Backbone Provenance in the collection level object than G1. One collection object could have more than one directed acyclic Backbone Provenance to other collection objects. This improvement is going to add a challenge to complicate collection level provenance in PID KI and Collection API. The data structure is shown in Fig. \ref{case2} with timeliness and provenance.

\textbf{\textit{G3. Directed acyclic backbone relationship among collection or member objects}}. This synthetic case is developed based on the case G2 but adds member-to-member provenance including [\textit{wasDerviedFrom, wasRevisionOf, wasQutoedFrom, hasPrimarySource, alternateOf and specializationOf}], which makes the provenance graph complex and challenges the PID KI and Collection API to distribute the collection and member level provenance.

G3 contains 100 collection objects, each collection connecting 4 member objects through the 'HadMember' relationship. Collection or Member objects could additionally be connected through [\textit{wasDerviedFrom, wasRevisionOf, wasQutoedFrom, hasPrimarySource, alternateOf and specializationOf}] to same objects of the same type. 

\begin{figure}[h]
\includegraphics[width=0.5\textwidth]{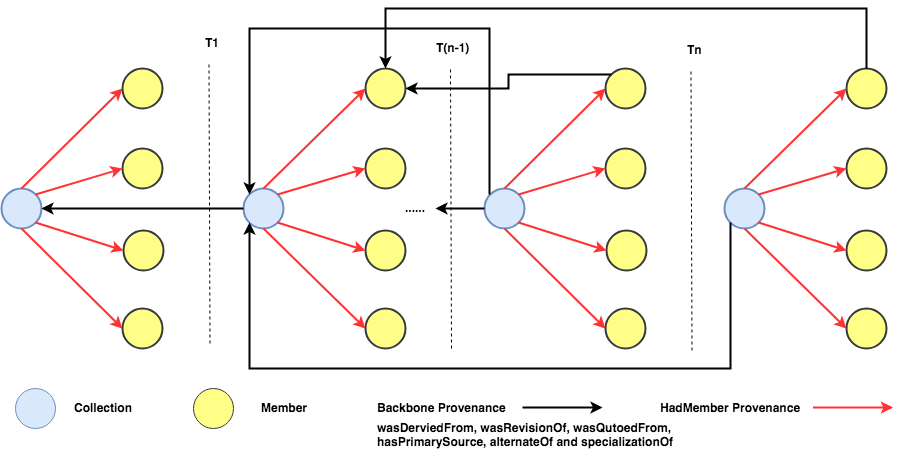}
\caption{Directed acyclic backbone relationship among collection or member objects}
\centering
\label{case3}
\end{figure}

In this case, one collection or member object can have more than one directed acyclic link to other collection or member objects. The data structure is shown in Fig. \ref{case3} exhibiting both timeliness and provenance relationship.

\begin{figure}[h]
\includegraphics[width=0.5\textwidth]{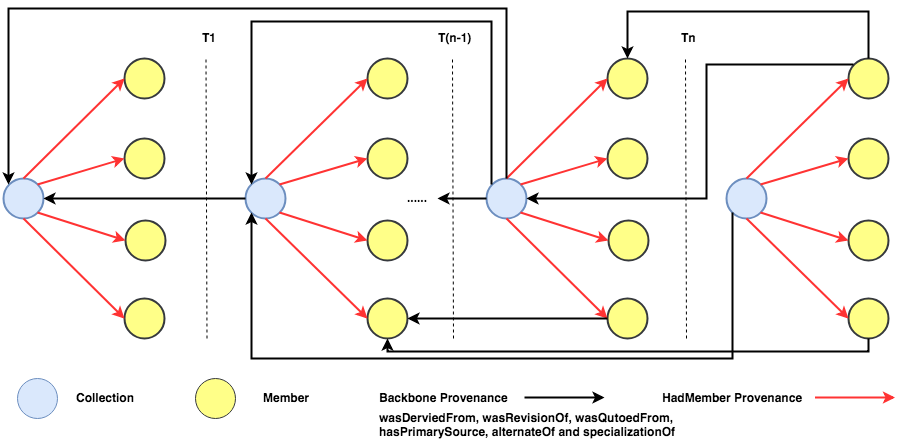}
\caption{Directed acyclic backbone relationship among all objects}
\centering
\label{case4}
\end{figure}

\textbf{\textit{G4. Directed acyclic backbone relationship among all objects}}. Directed acyclic Provenance relationship contains collection and member objects with directed acyclic Backbone Provenance among them. Still, one collection object connects its member objects by 'HadMember' relationship, and collection object and member object could connect to different type objects with Backbone Provenance in this provenance relationship graph. G4 has the greatest challenge for PID KI and Collection API to distribute provenance since unexpected provenance relationships exist everywhere. Also, while distributing provenance in PID KI and Collection API, both of them will face the challenge of efficiently presenting collection and member level provenance. The data structure is presented in Fig. \ref{case4}, showing the timeliness and provenance relationships in objects.

\section{Information Distribution}\label{sec:solution}

Handle System and Collection API both can saving provenance information in their objects. We study three different approaches to distribute the provenance information:

\begin{itemize}
    \item 
I1: Provenance maintained in the PID Kernel Information record.  HadMember is an invocation from the Collection API
\item
I2: The provenance is distributed in PID KI and Collection API.
\item
I3: Collection API maintains the provenance.
\end{itemize}

\textbf{\textit{I1: Provenance maintained in the PID Kernel Information record.  HadMember is an invocation from the Collections API}}

In this option, Backbone Provenance relationships are maintained as PID Kernel Information.  Only the 'HadMember' relationship is maintained within Collection API model. 
 
The PIDs are the vertices/nodes in the provenance relationship. By resolving PID KI, consumers are able to construct a large proportion of the provenance relationship. However, if two objects are related through the 'HadMember' provenance, the resolving client must query the Collection API to construct the remaining provenance relationship. The HadMember relationship is maintained in Collection-Member data model, pointing to the PIDs of relevant digital objects.

\textbf{\textit{I2: The provenance is distributed in PID KI and Collection API}}. Instead of saving the entire Backbone Provenance relationship within the PID KI record, this approach distributes provenance relationships into PID KI and the Collection-Member data model. The member object in Collection API is going to replace the PID KI by maintaining provenance linking to next PIDs. While resolving the Collection API objects, it will reduce one step of resolving the current PID KI of Collection API objects, and directs the process to the next level provenance.

\textbf{\textit{I3: Collection API maintains the Backbone Provenance}}. The strategy of \textit{I3} is simple, directly using the Collection-Member model of Collection API to maintain the provenance information. During the resolving process, a client accesses only a single PID to start the workload, and then must iteratively access the Collection API to retrieve the provenance information to construct the provenance graph. 

\section{Experimental Setup and Architecture}\label{sec:evaluation}

We proposed to evaluate the strength and weakness of PID KI and Collection API considering the proposed solutions and use cases in below section with proper workload and evaluation metrics.

We launch the test instance on Amazon AWS with:

\begin{table}[h!]
    \begin{adjustbox}{width=0.5\textwidth}
    \centering
    \begin{tabular}{|c|c|}
        \hline  
        Instance Provider & Amazon EC2 \\ \hline
        instance type & t2.xlarge \\ \hline
        CPU & 4 Core \\ \hline
        Memory & 16 GB \\ \hline
        system & Ubuntu, 16.04 LTS \\ \hline
        Handle system & version 9.0.4 \\ \hline
        Collection API & original version  \\ \hline
        Configuration for Handle and Collection API & default setting \\ \hline
        Handle Port & 8080 \\ \hline
        Collection Port & 5000 \\ \hline
    \end{tabular}
    \label{tab:my_label}
    \end{adjustbox}
\end{table}

In normal workflow, a client first acquires a PID KI with provenance information from the Handle System, then follows the provenance track to acquire the Collection or Member object with provenance information from Collection API. In some cases, a query request is made repeatedly to the same service over a period of time in order to obtain the same type objects based on the use case being tested. Also, at the same time, only one of the service, Handle system or Collection API service, is working for processing the requests, and the other one is listening.

\begin{figure}[h]
\includegraphics[width=0.5\textwidth]{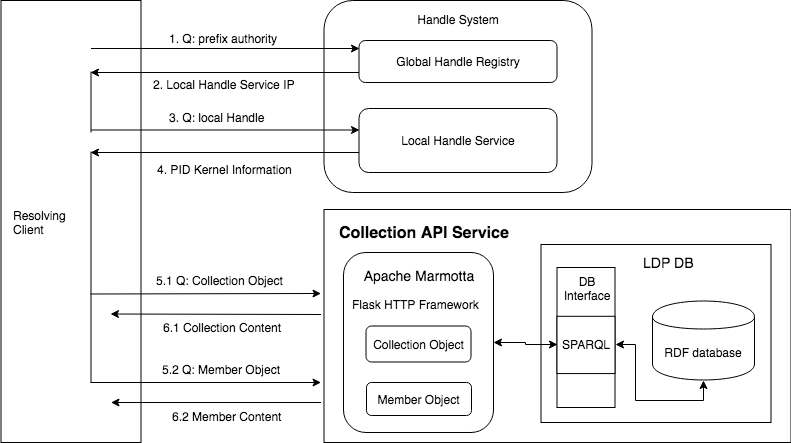}
\caption{Server Architecture with Handle System and Collection API}
\centering
\label{serverarchitecture}
\end{figure}

Architecture details of Collection API are presented in Fig. \ref{serverarchitecture}. Collection API is implemented in Python with Flask HTTP Framework. The request is passed from the HTTP interface to an RDF database though layered components. The DB interface maps the collection-member model into graph-based RDF entries through a SPARQL endpoint. 

\subsection{Workload Description}

In the experiment, we test the three information distributions $\{I1, I2, I3\}$ and four graphs $\{G1, G2, G3, G4\}$.  Each test has a unique workload with different algorithms to query services and build the provenance graph. In \textit{\textbf{I1}}, the graph would be built by acquiring PIDs with 'HadMember' related objects; in \textbf{\textit{I2}}, the graph would be built by acquiring PIDs and Member objects of Collection API with fine-grained relationships; in \textbf{\textit{I3}}, the graph would be built by acquiring Collection and Member objects from Collection API.

\textit{\textbf{Test 1: the graph would be built by acquiring PIDs with 'HadMember' related objects}}

The workload simulates the environment of Collection API and Handle System. With the pilot attempt for Collection API, we embed it with the light workload for affecting the performance of Handle record lightly. 

We assign PIDs for all digital objects, and each PID KI contains a block of provenance information with \{wasDerviedFrom, wasRevisionOf, wasQutoedFrom, hasPrimarySource, alternateOf and specializationOf\}. The Backbone Provenance is drawn by PIDs, and ‘hadMember’ provenance is saved within Collecion API objects.

\begin{figure}[]
\includegraphics[width=0.5\textwidth]{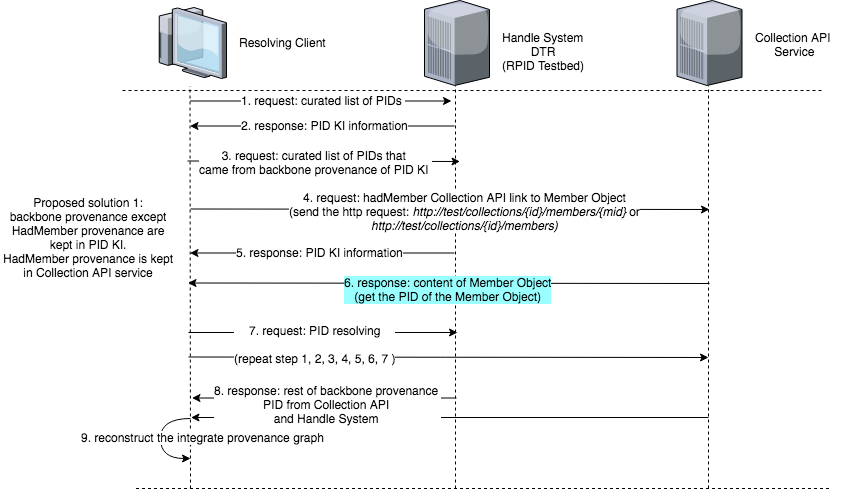}
\caption{PID Backbone relationship retrieving workload diagram}
\centering
\label{test1}
\end{figure}

The workload simulates the HTTP GET queries from resolving client to retrieve Backbone Provenance in PID KI from Handle System with a ‘hadMember’ link referencing to Collection API Member objects. The client issues another HTTP GET query to retrieve all member objects in Collection API, and then access the PID KI of member objects from Handle System to acquire Member objects' provenance. At this point, we obtain the Provenance relationship of one collection object. The client repeatedly issues the HTTP GET to traverse through the provenance relationship to reach the endpoints. The workload diagram is demonstrated in Fig. \ref{test1}.

\textbf{\textit{Test 2: the graph would be built by acquiring PIDs and provenance in the Collection API member objects}}

The Test 2 balances the workload between Handle System and Collection API. Comparing to Test 1, the Collection API takes more workloads in building the provenance relationship. Moving partial Backbone Provenance from PID KI to Member Object, drawing Backbone Provenance with both PIDs and Collection API member object, and skipping one step of retrieving PID KI for retrieved Member objects.

\begin{figure}[]
\includegraphics[width=0.5\textwidth]{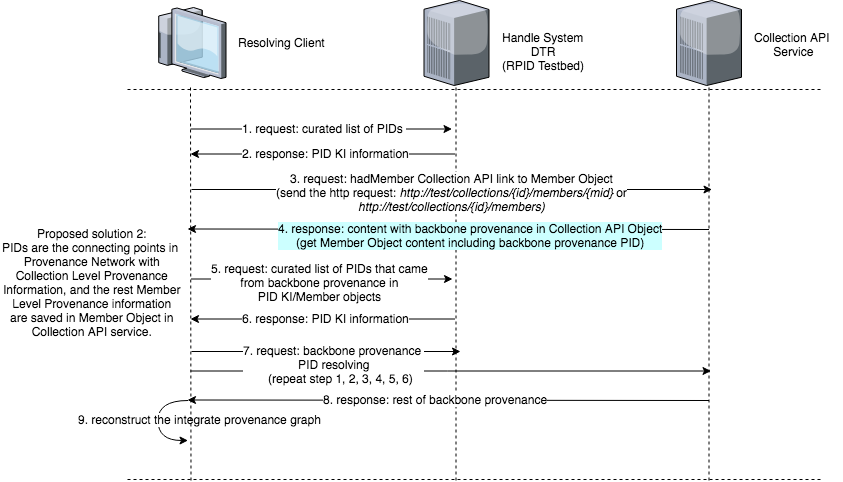}
\caption{PID and member object graph retrieving workload diagram}
\centering
\label{test2}
\end{figure}

The workload consists of HTTP GET queries from resolving client to retrieve PIDs and Collection API Member objects. Once the client acquires the content of Collection API member objects, it will find a block of provenance information with \{wasDerviedFrom, wasRevisionOf, wasQutoedFrom, hasPrimarySource, alternateOf and specializationOf\}, which is same as the provenance block in PID KI. All these provenance links within Member object provenance block are the references to the PID KI of the next node. By the time, HTTP GET queries have accessed the PID KI and Member object once to get the provenance relationship of one entire collection object with related objects. Further, the client traverses though the edges of relationship to reach the endpoints and draw the provenance relationship. The workload diagram is demonstrated in Fig. \ref{test2}.

\textbf{\textit{Test 3: the graph would be built by acquiring Collection and Member objects in Collection API}}

In the last exploration, the Collection API maintains all provenances instead of PID KI. Moving all Backbone Provenance from PID KI to Collection/Member Object, drawing Backbone Provenance relationship only within Collection API Member and Collection objects, and skipping the step of retrieving PID KI for any Collection API objects.

\begin{figure}[h]
\includegraphics[width=0.5\textwidth]{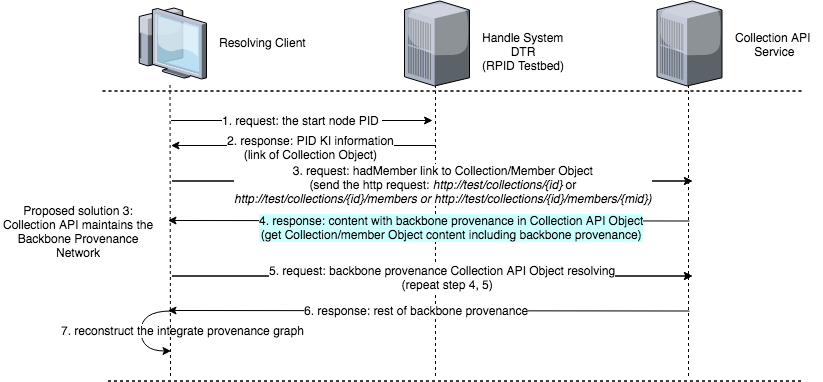}
\caption{collection and member object graph retrieving workload diagram}
\centering
\label{test3}
\end{figure}

HTTP GET queries are simulated by resolving client to retrieve digital objects in Collection API. Both Collection object and Member object contains a block of provenance information with \{wasDerviedFrom, wasRevisionOf, wasQutoedFrom, hasPrimarySource, alternateOf and specializationOf\}. The communications are issued within Collection API, building the provenances relationship by collection nodes and member nodes. The workload diagram is demonstrated in Fig. \ref{test3}.

\section{Experimental results and Analysis}\label{sec:result}
There are two kinds of experiments in this section: \textit{Baseline Experiment} and \textit{Non-baseline Experiment}. Before doing these two experiments, we also make a experiment for depositing the data into the environment and collect the metrics of throughput and cost.

\subsection{Depositing Data}
The preparation of the experiment is depositing digital objects into the Handle System and Collection API, and it also tests the registering performance for PID Object, Collection Object and Member Object. As we described above, we have four use cases. So for each solution, we would totally deposit 15000 objects, and more details are presented in table \ref{deposit}.

\restylefloat{table}
\begin{center}
\begin{table}[h]
\begin{adjustbox}{width=0.5\textwidth}
\begin{tabular}{|l|l|l|l|l|}
\hline
                  & G1    & G2    & G3  & G4    \\ \hline
PID Object        & 9000 objects & 9000 objects & 9000 objects & 9000 objects \\ \hline
Collection Object & 1500 objects & 1500 objects & 1500 objects & 1500 objects \\ \hline
Member Object     & 6000 objects & 6000 objects & 6000 objects & 6000 objects \\ \hline
\end{tabular}
\end{adjustbox}
\caption{Numbers of digital objects deposited into Collection API and Handle System}
\label{deposit}
\end{table}
\end{center}

\begin{figure}[]
\includegraphics[width=0.45\textwidth]{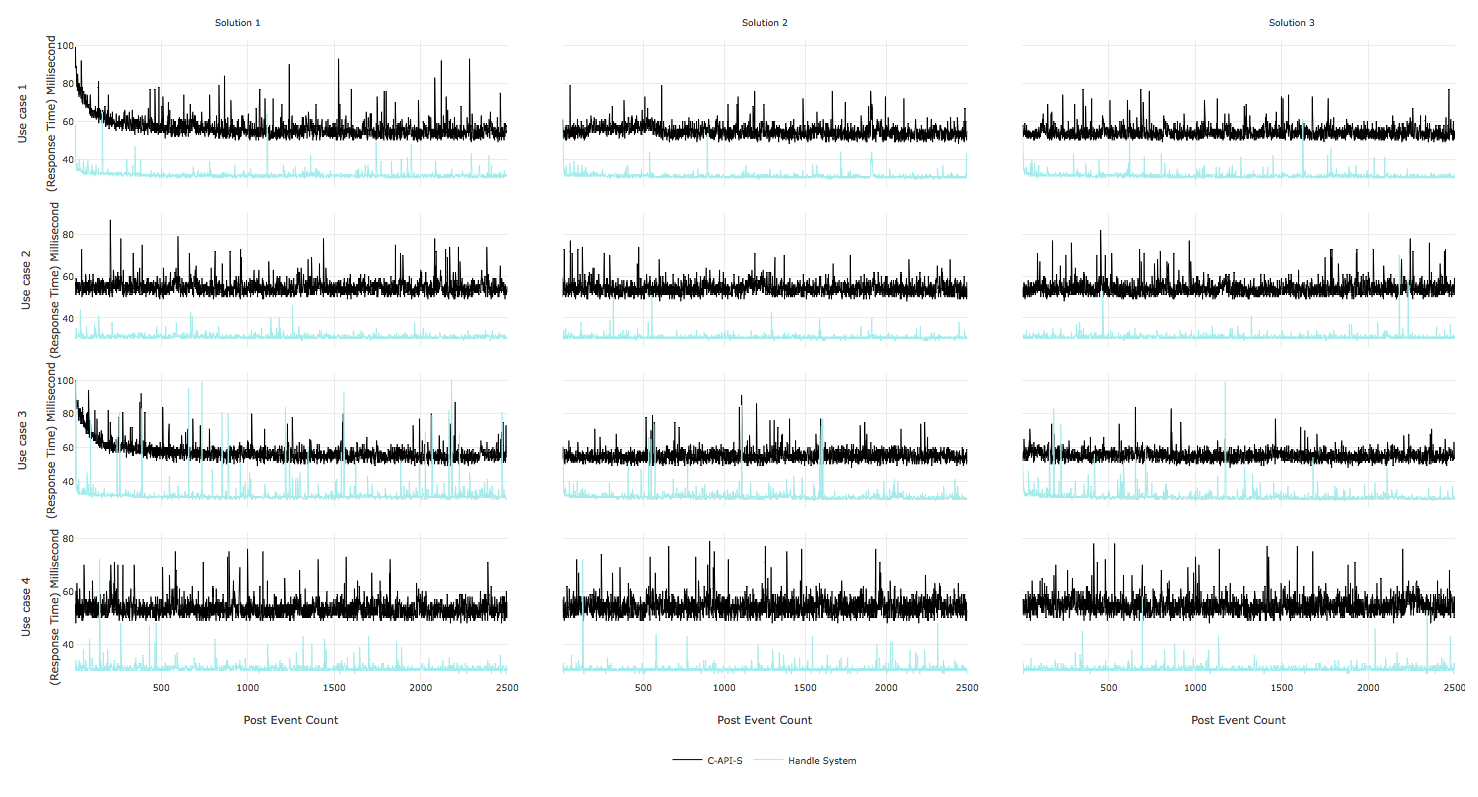}
\caption{Registration Cost of Handle objects and Collection API objects in Client}
\centering
\label{registerinclient}
\end{figure}

All these digital objects are registered from client side by using three solution approaches. 66000 registration requests (36000 PID objects, 6000 Collection objects, 24000 Member Objects) are sequentially sent by the client. Meanwhile, the client records the cost of entire registration of each request. The registration costs are listed in the Fig. \ref{registerinclient}. Comparing the registration cost of Handle System and Collection API from client side, we observe that the Handle System consumes around 31 milliseconds and Collection API consumes around 54 milliseconds to register the objects.

\begin{figure}[]
\includegraphics[width=0.45\textwidth]{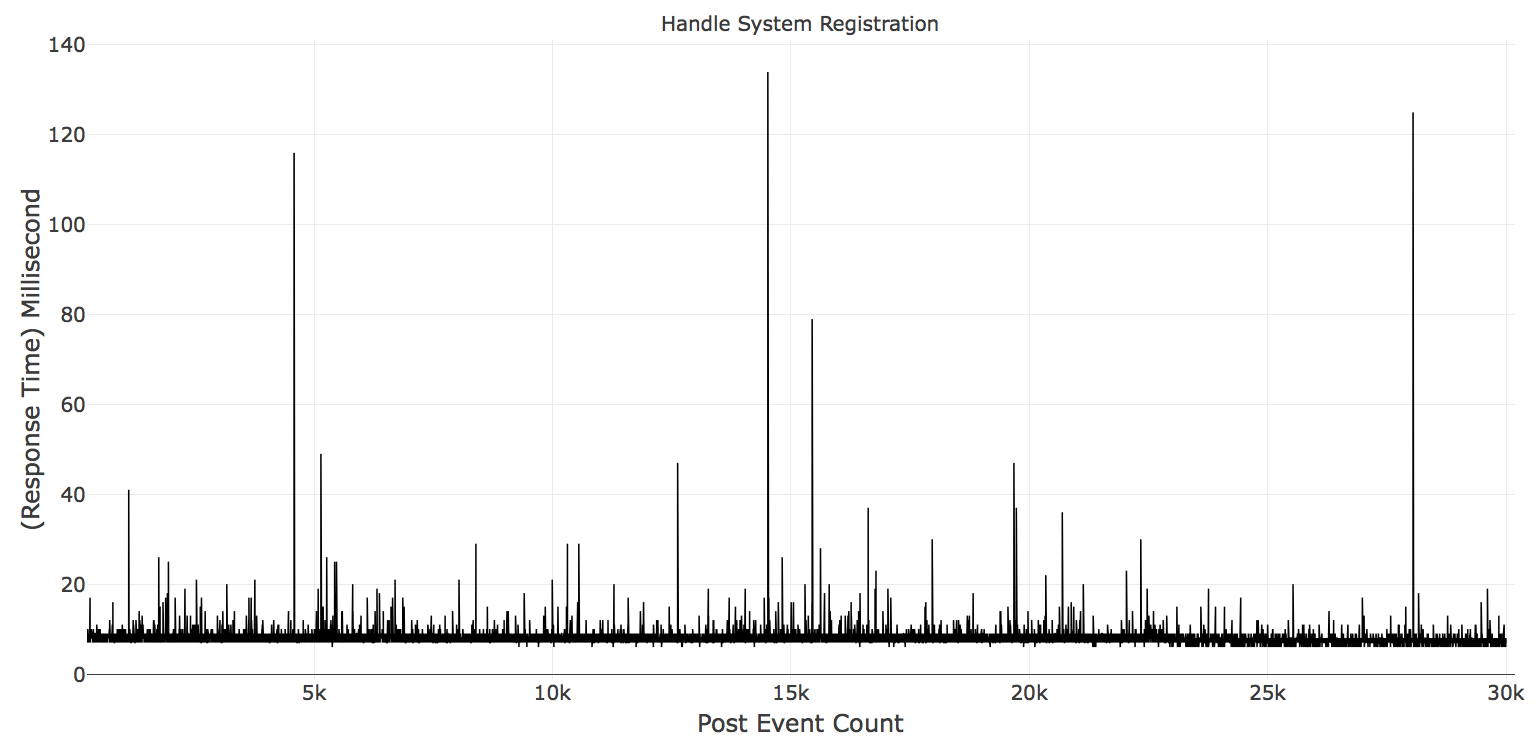}
\includegraphics[width=0.45\textwidth]{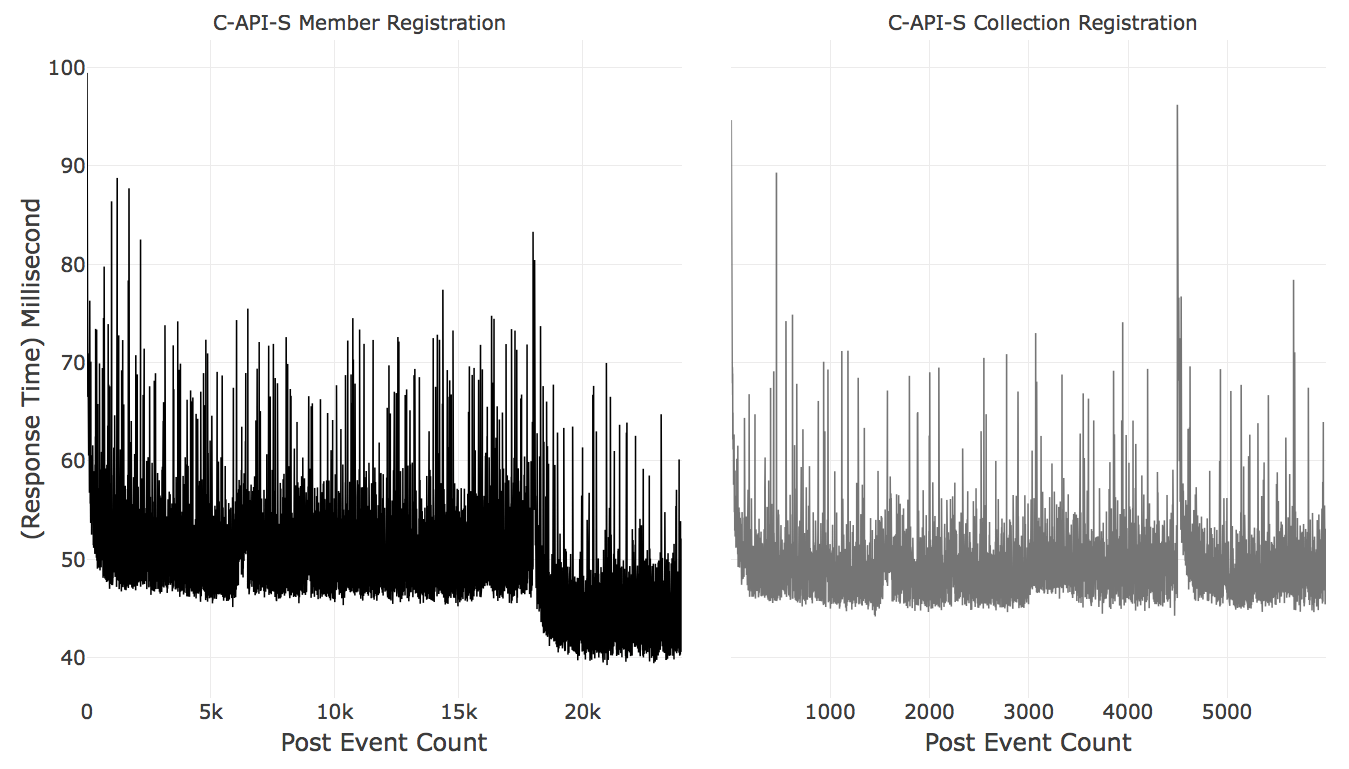}
\caption{Registration Cost in Handle System and Collection API}
\centering
\label{registerinserver}
\end{figure}

In the Server side, we also observe the internal cost metrics of registering an object in Handle System and Collection API, shown in Fig. \ref{registerinserver}. The Collection API costs 49.6 milliseconds on Collection Objects registration and 49 milliseconds on Member objects registration. The Handle System costs 8 milliseconds on registering the objects.

In registering process, Handle system has an obvious advantage of registering PIDs for objects.

\subsection{Resolving Algorithm for Handle Object and Collection API Object}
In the client, it will send HTTP request to two interface: the HTTP Get request is sent to REST API Servlet of Handle System, calling Berkeley Database to get the handle value; the HTTP Get request is sent to the HTTP Interface of Collection API, calling RDF database (provided by Apache Marmotta) to get the Collection/Member values. After resolving, the client acquires the values from Handle System and Collection API, it will extract the provenance block, resolving the PID or Collection API URL to relative services recursively. The recursive algorithm of traversing the backward Provence relationship started from the original PID is presented in Algorithm. \ref{algorithm}. 

\begin{algorithm} [h]
        \caption{\textbf{:} Resolving Algorithm for Handle objects and Collection API objects}
        \label{algorithm}
        
\begin{flushleft}
        \textbf{input:} Original PID of the Handle Object\\
        \textbf{output:} The Provenance relationship of Original PID
\end{flushleft}
\begin{algorithmic}[1]
\State $URL \gets OriginalPID_{0};$
\Function{GetProvrelationship}{URL}
\State $ID \gets Empty;$
\State $Provenance \gets Empty;$

\If{$isPID(URL)$}
    \State $content \gets callHandle(URL);$
    \State $ID \gets getID(content);$
    \State $Provenance \gets getProv(content);$
\Else
    \State $content \gets callCOLLAPI(URL);$
    \State $ID \gets getID(content);$
    \State $Provenance \gets getProv(content);$
\EndIf
\For{$\/{prov:value\/} \in Provenance$}
    \If{$!isEmpty(value)$}
        \State $setProrelationship(ID,prov,value);$
        \State $GetProvrelationship(value);$
    \EndIf
\EndFor
\EndFunction
\end{algorithmic}
\end{algorithm}

\subsection{Baseline Experiment}
In the \textit{Baseline Experiment} section and \textit{None-Baseline Experiment} section, we are doing the experiments on resolving the digital objects from Collection API and Handle System based on the different distribution approaches. The purpose of representing the data is evaluating the strength and weakness of distributing data provenance in PID KI and Collection API.

In this section, we observe the workload and performance of Collection API and Handle System. While querying objects in both services, each of them would have different algorithms to process HTTP GET queries. Therefore, we pay attentions to how they respond on these HTTP GET queries. 

\subsubsection{Handle System Querying}
The Handle System processing the HTTP GET querying straightly since there is only one type of Handle value, key-value paired PID Kernel Information. While REST API Servlet of Handle System receiving the HTTP GET request, it will first parse the Handle (PID) from the request, double checking the authorization information, then sends the query of Handle to the Berkeley database, and finally returns the JSON formatted result to consumers. In this experiment, the only cost is generated in accessing database and transferring the data.

\subsubsection{Collection API Querying}
Collection API chooses a little bite complex algorithm to deal with Collection and Member objects query requests. The HTTP interface receives the HTTP GET Query of Collection or Member objects, parsing the required information, then building the select query in SPARQL endpoints, and then query the object in RDF database.

\subsubsection{Querying Workload in Server}
\begin{figure}[h]
\includegraphics[width=0.5\textwidth]{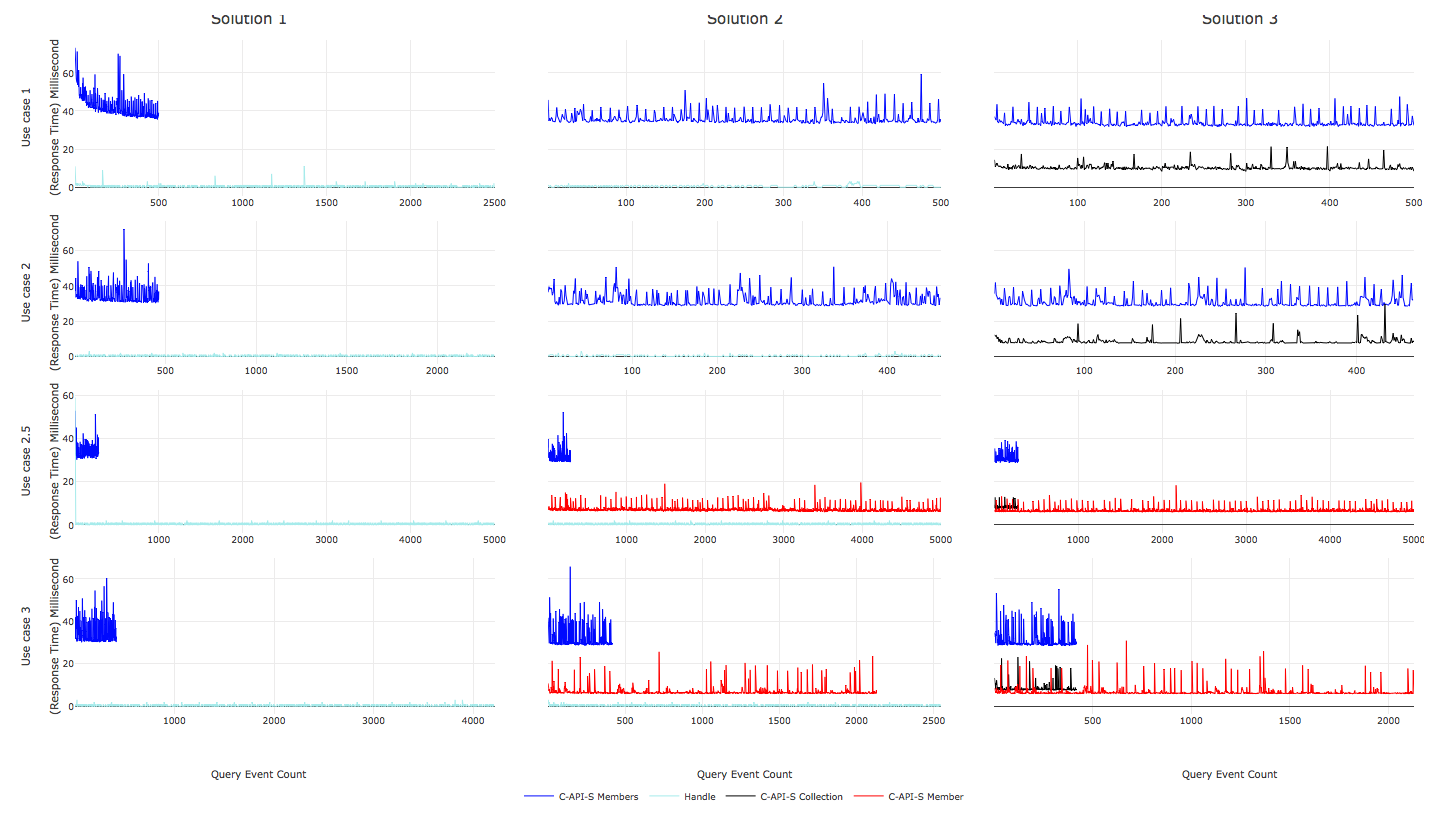}
\caption{Resolving Cost of Handle objects and Collection API objects in Server}
\centering
\label{serverresolve}
\end{figure}

We simulate some synthetic use cases as real world data, and the querying workloads of G2, G3 and G4 cannot access all objects since we randomly assign the provenance relationship among objects. Totally, we send 85730 requests to the Handle System and Collection API, excluding the network latency, observing the internal cost of processing queries in both servers, more details are presented in Table. \ref{serversolvesummary} and Fig. \ref{serverresolve}

\restylefloat{table}
\begin{table*}[h]
\begin{adjustbox}{width=1\textwidth}
\begin{tabular}{c|c|c|c|c|c|c|c|c|c|}
\cline{2-10}
                                                 & \multicolumn{3}{c|}{I 1}                 & \multicolumn{3}{c|}{I 2}           & \multicolumn{3}{c|}{I 3}          \\ \hline
\multicolumn{1}{|c|}{\multirow{4}{*}{G1}} &              & Members         & Handle         & Members     & Member       & Handle       & Collection  & Members     & Member       \\ \cline{2-10} 
\multicolumn{1}{|c|}{}                           & Volume         & 500 requests     & 2500 requests   & 500 requests & N/A          & 500 requests  & 500 requests & 500 requests & N/A          \\ \cline{2-10} 
\multicolumn{1}{|c|}{}                           & average cost & 42.12 ms        & 0.38 ms        & 35.84 ms    & N/A          & 0.62 ms      & 10.32 ms    & 33.96 ms    & N/A          \\ \cline{2-10} 
\multicolumn{1}{|c|}{}                           & average total        & \multicolumn{2}{c|}{22010 ms}    & \multicolumn{3}{c|}{18230 ms}             & \multicolumn{3}{c|}{22140 ms}            \\ \hline
\multicolumn{1}{|c|}{\multirow{3}{*}{G2}} & Volume         & 463 requests      & 2315 requests   & 463 requests & N/A          & 463 requests  & 463 requests & 463 requests & N/A          \\ \cline{2-10} 
\multicolumn{1}{|c|}{}                           & average cost & 34.39 ms        & 0.2 ms         & 32.01 ms    & N/A          & 0.33 ms      & 8.55 ms     & 31.08 ms    & N/A          \\ \cline{2-10} 
\multicolumn{1}{|c|}{}                           & average total        & \multicolumn{2}{c|}{16385.57 ms} & \multicolumn{3}{c|}{14973.42 ms}          & \multicolumn{3}{c|}{18348.69 ms}         \\ \hline
\multicolumn{1}{|c|}{\multirow{3}{*}{G3}} & Volume         & 287 requests      & 16571 requests   & 287 requests & 15136 requests          & 15423 requests  & 287 requests & 287 requests & 15136 requests         \\ \cline{2-10} 
\multicolumn{1}{|c|}{}                           & average cost & 33.46 ms        & 0.18 ms        & 31.2 ms    & 6.98 ms          & 0.19 ms      & 8.28 ms    & 30.79 ms     & 6.51 ms        \\ \cline{2-10} 
\multicolumn{1}{|c|}{}                           & average total        & \multicolumn{2}{c|}{12585.8 ms}  & \multicolumn{3}{c|}{117534.1 ms}          & \multicolumn{3}{c|}{109748.4 ms}         \\ \hline
\multicolumn{1}{|c|}{\multirow{3}{*}{G4}} & Volume         & 417 requests     & 4214 requests   & 417 requests & 2129 requests & 2546 requests & 417 requests & 417 requests & 2129 requests \\ \cline{2-10} 
\multicolumn{1}{|c|}{}                           & average cost & 33.78 ms        & 0.17 ms        & 31.48 ms    & 6.63 ms      & 0.18 ms      & 8.35 ms     & 30.92 ms    & 6.57 ms      \\ \cline{2-10} 
\multicolumn{1}{|c|}{}                           & average total        & \multicolumn{2}{c|}{14802.64 ms} & \multicolumn{3}{c|}{27700.71 ms}          & \multicolumn{3}{c|}{30363.12 ms}         \\ \hline
\end{tabular}
\end{adjustbox}
\caption{Summary of Resolving workloads in Server}
\label{serversolvesummary}
\end{table*}

In the Table. \ref{serversolvesummary}, we list four types of request: \textit{Members, Member, Collection and Handle}. \textit{Members} is a Collection API query request to get all member objects of a collection. \textit{Member} is a Collection API query request to get 1 member object of a collection. \textit{Collection} is a Collection API query request to get 1 collection. And \textit{Handle} is a Handle System query request to get 1 PID KI record.

The analysis of the experiment could be started on same case in different solution approaches. In this runtime metric, we got 12 results based on 4 cases and 3 solution approaches. For each case, we apply 3 solution approaches on distributing provenance in Handle System and Collection API. 

Overall, the Collection API cost on resolving is much higher than the Handle System. The average Handle resolving cost is less than 1 ms. And resolving cost of any objects in Collection API is more than 1 ms. 

\textbf{\textit{G1: Directed linear backbone relationship among collection objects:}}

G1 presents the simplest relationships among objects. The first concern is the average total cost of different solutions on representing the G1 provenance. The solution 2 cost the least time on all three solutions. The advantage of solution 2 is less Handle queries in its entire workloads because we distribute the provenance into Collection API member objects. Solution 1 processes 2500 Handle objects for Collection API objects, 500 handles for collection objects and 500 $\times$ 4 handles for member objects, which is more than solution 2 (500 handles). Even though solution 3 processes 500 Collection objects in Collection API instead of PID KI in Handle system, querying the Collection object in Collection API needs more time than PID KI object.

A early conclusion for this subsection is that PID KI is the best choice for presenting single and simple digital object with backbone provenance relationship.

\textbf{\textit{G2: Directed Acyclic backbone relationship among collection objects:}}

Analyzing the result of the three solution approaches for the G2, we get the same conclusion that PID KI is the best choice to present the single object. Service will take huge advantage on resolving the object within Handle System resulted from its good architecture and database design. 

Further, both G1 and G2 show the I2 is the best choice to handle the graph cases because of less PID KI requests. Considering the solution 2 strategy, it puts the provenance relationships in Member object within Collection API model instead of using PID KI for that Member object. When we resolve one Collection API member object, we could get all provenance information from it, there is no need to go through the PID KI object. For example, if one Member object has no provenance relationship, we can know it and don not query PID KI object; Otherwise, we could directly resolve the next PIDs. 

In these two subsections, we only set the provenance relationship in collection-to-collection level objects, so the result is similar between the experiment of G1 and G2. In the next subsection, we are testing the case adding member-to-member Backbone Provenance relationship to observe more information.

\textbf{\textit{G3: Directed acyclic backbone relationship among collection or member objects:}}

The G3 shows different result since we make more complex data structure than G2. 

In the result, we observe that the average cost of querying 1 Member object is less than querying 1 Collection object, and 4 times the cost of 1 Member object querying is still less than cost of querying all Member objects. We assume the cost of 1 Member object querying should be equal to the average cost of 4 Member objects querying. Actually, it is not because of the different query algorithms. All Member objects querying would take more time on database querying rather than 4 times the cost of 1 Member querying, especially on data existing check and number of database accessing. 

In this experiment, we conclude that one query of all Member object is not appropriate but the only approach to handle a fine-gain provenance relationship between Collection and Member objects.

The G3 helps the client acquire a clear understanding of the Collection API's weakness and strength: 1) performance order of functionalists to represent objects is: \textbf{1 Member} $>$ \textbf{1 Collection} $>$ \textbf{All Members}. 2) All of three query functionalists in Collection API are less efficient than PID querying. 3) Collection API has three types of digital object and could maintain the fine-grain relationships for objects than PID KI. 

Overall, G3 gives us a different conclusion that I1 maximizes the advantage of PID KI. I2 and I3 spend too much time on presenting the data structure in Collection API. 

\textbf{\textit{G4: Directed acyclic backbone relationship among allobjects}}

The G4 simulates the real world case, applying the very complex provenance relationships in data structure. I1 is the best solution resulted from the strategy of distributing provenance between PID KI and Collection API model. Comparing with I1, I2 and I3, 91\% of requests in I1 are resolving the PID KI, which saves a lot of time in entire resolving process. Even though I3 has the smallest amount of resolving requests, the Collection API resolving request cost more time than PID KI on resolving the same amount of provenance. This use case magnifies the efficiency of maintaining provenance in PID KI.

\subsection{Non-baseline Experiment}
\begin{figure}[h]
\includegraphics[width=0.5\textwidth]{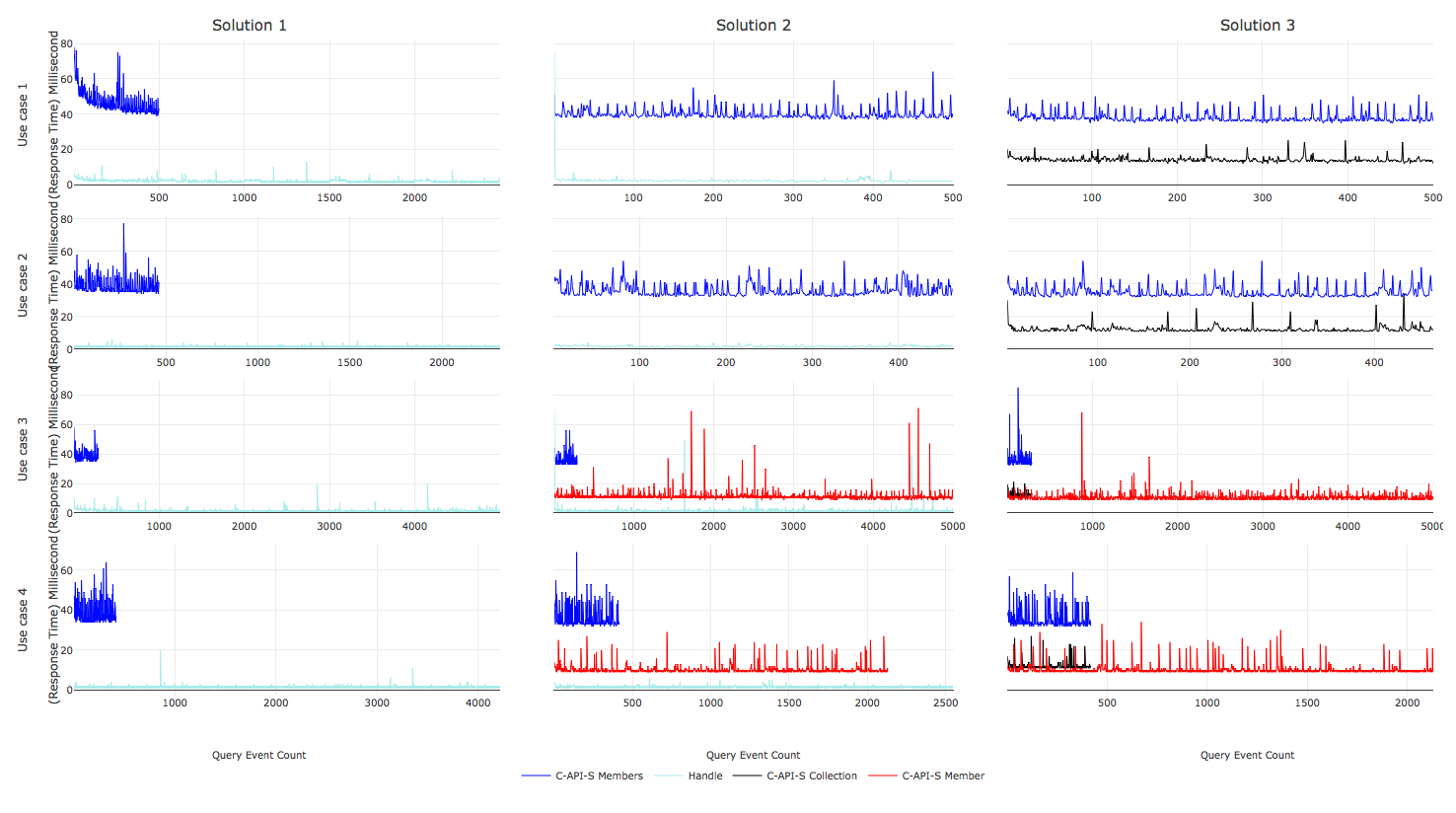}
\caption{Resolving Cost of Handle objects and Collection API objects in Client}
\centering
\label{clientresolve}
\end{figure}

Addition to the \textit{Baseline Experiment}, we observe the runtime metrics from client, caring about how to represent the provenance relationship for consumers. The Fig. \ref{clientresolve} shows the workloads in each case and its result. The Table. \ref{clientresolvesummary} summaries the resolving work for all workloads.

\restylefloat{table}
\begin{table*}[h]
\begin{adjustbox}{width=1\textwidth}
\begin{tabular}{c|c|c|c|c|c|c|c|c|c|}
\cline{2-10}
                                                 & \multicolumn{3}{c|}{I1}                 & \multicolumn{3}{c|}{I2}           & \multicolumn{3}{c|}{I3}          \\ \hline
\multicolumn{1}{|c|}{\multirow{4}{*}{G1}} &              & Members         & Handle         & Members     & Member       & Handle       & Collection  & Members     & Member       \\ \cline{2-10} 
\multicolumn{1}{|c|}{}                           & Volume         & 500 requests     & 2500 requests   & 500 requests & N/A          & 500 requests  & 500 requests & 500 requests & N/A          \\ \cline{2-10} 
\multicolumn{1}{|c|}{}                           & average cost & 46.21 ms        & 1.94 ms        & 39.94 ms    & N/A          & 2.33 ms      & 14.1 ms     & 37.89 ms    & N/A          \\ \cline{2-10} 
\multicolumn{1}{|c|}{}                           & total        & \multicolumn{2}{c|}{27955 ms}    & \multicolumn{3}{c|}{21135 ms}              & \multicolumn{3}{c|}{25995 ms}            \\ \hline
\multicolumn{1}{|c|}{\multirow{3}{*}{G2}} & Volume         & 463 object      & 2315 requests   & 463 requests & N/A          & 463 requests  & 463 requests & 463 requests & N/A          \\ \cline{2-10} 
\multicolumn{1}{|c|}{}                           & average cost & 38.35 ms        & 1.63 ms        & 35.97 ms    & N/A          & 1.79 ms      & 12.25 ms    & 34.9 ms     & N/A          \\ \cline{2-10} 
\multicolumn{1}{|c|}{}                           & total        & \multicolumn{2}{c|}{21529.5 ms}  & \multicolumn{3}{c|}{17399.54 ms}          & \multicolumn{3}{c|}{21830.45 ms}         \\ \hline
\multicolumn{1}{|c|}{\multirow{3}{*}{G3}} & Volume         & 287 object      & 16571 requests   & 287 requests & 15136 requests          & 15423 requests  & 287 requests & 287 requests & 15136 requests         \\ \cline{2-10} 
\multicolumn{1}{|c|}{}                           & average cost & 37.88 ms        & 1.65 ms        & 35.37 ms    & 10.84          & 1.7 ms      & 12.22 ms    & 35.35 ms     & 10.25         \\ \cline{2-10} 
\multicolumn{1}{|c|}{}                           & total        & \multicolumn{2}{c|}{38213.71 ms}  & \multicolumn{3}{c|}{277940.2 ms}          & \multicolumn{3}{c|}{168796.6 ms}         \\ \hline
\multicolumn{1}{|c|}{\multirow{3}{*}{G4}} & Volume         & 417 requests     & 4214 requests   & 417 requests & 2129 requests & 2546 requests & 417 requests & 417 requests & 2129 requests \\ \cline{2-10} 
\multicolumn{1}{|c|}{}                           & average cost & 37.79 ms        & 1.53 ms        & 35.3 ms     & 10.11 ms     & 1.592 ms     & 11.93 ms    & 34.48 ms    & 10.07 ms     \\ \cline{2-10} 
\multicolumn{1}{|c|}{}                           & total        & \multicolumn{2}{c|}{22205.85 ms} & \multicolumn{3}{c|}{40297.52 ms}          & \multicolumn{3}{c|}{40792 ms}            \\ \hline
\end{tabular}
\end{adjustbox}
\caption{Summary of Resolving workloads in Client}
\label{clientresolvesummary}
\end{table*}

The resolving result from client presents the communicating cost between server and users except the network latency. The results of 12 small experiments of \textbf{\textit{Non-baseline Experiment}}, 3 solutions $\times$ 4 use cases, is not different from the \textbf{\textit{Base-line Experiment}} we observed, what we concern here is the communication cost of transferring different types of objects. Handle object and Collection API object are different types with different sizes, especially the Collection object and Member object in Collection API, so the resolving process of them would cost different time on receiving the output from server to client. We conclude the communication cost of this experiment on Table \ref{communicationcost}. 

\restylefloat{table}
\begin{table*}[h]
\begin{adjustbox}{width=1\textwidth}
\begin{tabular}{cc|c|c|c|c|c|c|c|c|}
\cline{3-10}
                                  &                            & \multicolumn{2}{c|}{I1} & \multicolumn{3}{c|}{I2} & \multicolumn{3}{c|}{I3} \\ \cline{2-10} 
\multicolumn{1}{c|}{}             &                            & Handle         & Members        & Handle    & Members  & Member   & Collection  & Members & Member  \\ \hline
\multicolumn{1}{|c|}{G1}   & Average Communication Cost & 1.56 ms        & 4.09 ms        & 1.71 ms   & 4.1 ms   & N/A      & 3.78 ms     & 3.93 ms & N/A     \\ \hline
\multicolumn{1}{|c|}{G2}   & Average Communication Cost & 1.43 ms        & 3.96 ms        & 1.46 ms   & 3.96 ms  & N/A      & 3.7 ms      & 3.82 ms & N/A     \\ \hline
\multicolumn{1}{|c|}{G3} & Average Communication Cost & 1.47 ms        & 4.42 ms        & 1.51 ms   & 4.17 ms  & 3.86 ms  & 3.94 ms     & 4.56 ms & 3.74 ms \\ \hline
\multicolumn{1}{|c|}{G4}   & Average Communication Cost & 1.36 ms        & 4.01 ms        & 1.41 ms   & 3.82 ms  & 3.48 ms  & 3.58 ms        & 3.56 ms & 3.5 ms  \\ \hline
\end{tabular}
\end{adjustbox}
\caption{Summary of Network Communication Cost from Server to Client}
\label{communicationcost}
\end{table*}

We suppose the size of different type digital objects would not have significant effects on the communication cost, since in the experiment, we only put few bytes data (backbone provenance links) into the PID KI and Collection API object. The Handle System takes 1.985 milliseconds to transfer the data to client. Collection API takes 3.75 milliseconds to send the Collection object to client. However, the communication costs between All Members (4 Member objects) and 1 Member are similar, transferring All Members (4 Member objects) to users costs 4.04 milliseconds and transferring 1 Member to users costs 3.645 milliseconds. Hence, we can finally conclude that the size of Collection API object would not affect the communication performance of Collection API.

%\subsubsection{Extra Functionality in Collection API: GetProperty}

%In Collection API, the server provides other flexible functionalists, like joining two collection objects, finding intersection of two collection objects, finding matched collection objects, and returning property of member objects. Here, we do the special experiment on use case 3 to test the getting property of Member objects function. The getting property functions uses the same algorithm of querying Member object, reducing the content on transferring. The Fig. \ref{property} displays the Query Cost from Collection API server and client. The server side average cost is 6.44 milliseconds, the client side average cost is 9.95 milliseconds, and the average communication cost is 3.51 milliseconds.

%\begin{figure}[h]
%\includegraphics[width=0.5\textwidth]{property.png}
%\caption{Resolving Cost of Getting Property from Collection API}
%\centering
%\label{property}
%\end{figure}

\subsubsection{Extra: Why Collection API costs more time}

\begin{figure}[h]
\includegraphics[width=0.5\textwidth]{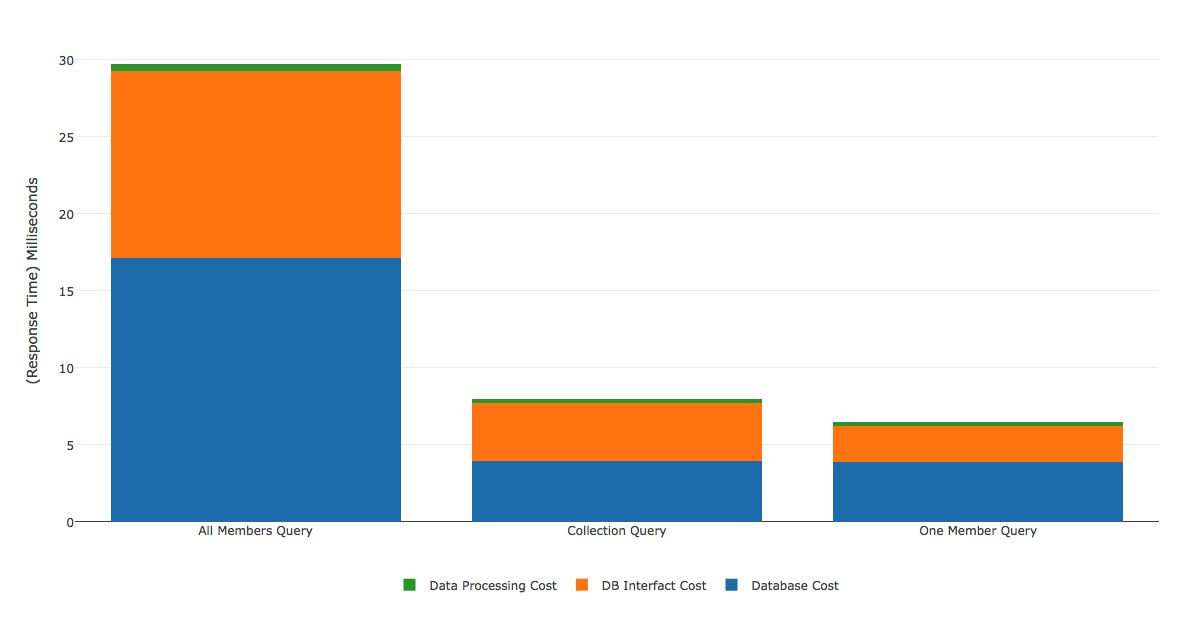}
\caption{Time Cost when Querying Collection API}
\centering
\label{timecost}
\end{figure}

The time cost of querying Collection API insistents of two main parts: database processing and DB Interface processing. The Database processing is ingesting data from RDF database, and DB Interface processing is converting data type from graph to Collection API object. The Fig. \ref{timecost} highly indicates the time cost in these two layers, and here is the reason why Collection API cost more time than Handle System. 

\section{Discussions and Summary}\label{sec:discussion}

This study is agnostic to the makeup of the Complex digital object or how it is managed.  It evaluates the pros and cons of information distribution between PID Kernel Information and Collection API, specifically Collection API. It illustrates how the two can interact to manage and represent the provenance relationship of Complex digital object. 

In this pilot evaluation, we simulate four use cases with three proposed solution approaches to test the different distributions of backbone provenance in the PID Kernel Information and Collection API. 

In PID KI, we aim to keep minimal kernel information free of complex contents, like lists of member information. This keeps the PID KI record small, and less susceptible to frequent change.  

Collection API presents a fine-grain relationship of objects through the Collection-Member data model. The Collection-Member model provides an easy way to handle the 'HadMember' relationship without violating the definition of PID Kernel Information. However, the Collection-Member data model of Collection API has some disadvantages. Collection API gains the benefit from Collection-Member model, presenting a clear graph of objects, meanwhile paying the cost of more time on data managing and querying. We rely on Collection API to help PID KI deal with complex relationships at minimal cost.

The experimental evaluation that we conduct tell us more information about the collaboration work between PID KI and Collection API, showing the undiscovered phenomenons, helping us make conclusions to manage and distribute the provenance relationship of Complex digital objects. 

Based on the use case we designed, we could classify the Provenance relationship into two categories: type-to-type and all-to-all. The type-to-type means only same type object would connect to each other, collection object to collection object and member object to member object. The all-to-all means all objects could connect to others, no matter collection object or member object, like combining files from different folders. These two categories could be a good summary that used to understand the use case and gain insights from the experiments that we did. 

\textbf{\textit{Information Distribution 1}} is the first attempt to add Collection API data structure into provenance relationship of PID KI, using the Collection-Member relationships to represent the 'HadMember' provenance. Since the PID KI maintains the Backbone provenance, and has less capacity to maintain complex provenance, like HadMember. Collection API only provides the information about the 'HadMember' relationship. This solution is a simple provenance distribution to represent the Complex digital objects, but showing a incredible output on use case simulations. 

\textbf{\textit{Information Distribution 2}} is a advanced way to use PID KI and Collection API, experimentally balancing the provenance information, moving partial provenance information from PID KI into Collection API. There is a trade-off between number of requests and time costs in Solution 2 strategy. For the simple provenance relationship, collection-level provenance relationship (type-to-type), solution saves the time by ignoring the redundant accesses to the PID KI, proved in G1 and G2. However, when the member-level (type-to-type) provenance relationship is embedded, Collection API consumes more time since querying performance of Collection API is lower than PID KI. We can conclude that the solution 2 is a alternative approach to solve the situation that solution 1 cannot handle efficiently. 

There is another conclusion that the volume of resolving requests and the percentage of PID KI in resolving requests are the main parameters to determine the better provenance distribution strategy. In G1 and G2, I2 is the best solution because it has the smaller volume than I1 and has a higher percentage of PID KI than I3. In G3 and G4, I1 is the best solution because it has the smaller volume than I2 and has a higher percentage of PID KI than I3.

\textbf{\textit{Information Distribution 3}} is a control group observing the Collection API's work on handling Provenance relationship individually. All experiment results show the pure Collection API is good at managing Provenance relationship, but is not efficient enough. PID KI and Collection API work together for mutual benefit, handling complex provenance relationship, and cost less time on representing the Complex digital object.

Also, for above three discussions, we add another conclusion about relationship communication cost of the Collection API and PID KI. There is no doubt about the advantage of PID KI, but for designing the data structure in Collection API, we encourage to use Collection and All Member operation since they both are efficient, saving time on representing and transferring the digital object.

\section{Conclusion}\label{sec:conclusion}
This pilot evaluation suggests that the Collections API approach offers significant potential for managing and representing collection digital objects through its flexible Collection-Member data model. Moreover, the provenance distribution in PID KI and Collection API follows two purposes: maximally utilizing the PID KI and reducing the redundancy of provenance resolving. Keeping the minimal size of objects in maintaining provenance, and maximally use the PID KI to maintain provenance. 

\section{Acknowledgements}
This work is funded in part by NSF grants \# 1659310 and \# 1839013.

{\footnotesize \bibliographystyle{IEEEtran}
\bibliography{bibliography}}

\end{document}